\documentclass[letterpaper,10pt,conference,english]{ieeeconf}
\IEEEoverridecommandlockouts
\usepackage[cmex10]{amsmath}
\usepackage[noadjust]{cite}
\usepackage{%
    amssymb,%
    babel,%
    color,%
    enumerate,%
    etoolbox,%
    mathtools,%
    stmaryrd%
}

\usepackage[ruled,vlined]{algorithm2e}
\usepackage{authblk}

\newcommand{\CE}{\color{black}} 
\begin{document}

\title{Deep Reinforcement Learning Based Power control for Wireless Multicast Systems}
\author[1]{Ramkumar Raghu} 
\author[1]{Pratheek Upadhyaya} 
\author[1]{Mahadesh Panju} 
\author[1,2]{Vaneet Aggarwal} 
\author[1]{Vinod Sharma}  
\affil[1]{Indian Institute of Science, Bangalore, INDIA. \textit{\{mahadesh,ramkumar,vinod\}@iisc.ac.in}}
\affil[2]{Purdue University, West Lafayette IN, USA.  \textit{vaneet@purdue.edu}}

\maketitle
\begin{abstract} 
We consider a multicast scheme recently proposed for a wireless downlink in \cite{wcnc}. It was shown earlier that power control can significantly improve its performance. However for this system, obtaining optimal power control is intractable because of a very large state space. Therefore in this paper we use deep reinforcement learning where we use function approximation of the Q-function via a deep neural network. We show that optimal power control can be learnt for reasonably large systems via this approach. The average power constraint is ensured via a Lagrange multiplier, which is  also learnt. Finally, we demonstrate that a slight modification of the learning algorithm allows the optimal control to track the time varying system statistics.
 
\end{abstract}
\section{Introduction}

Wireless networks are being constantly refined to cater for seamless delivery of huge amount of data to the end users. With increased user generated contents and proliferation of social networking sites, almost $78\%$ of mobile data traffic is expected to be due to mobile videos \cite{CVNI}. Also, the requested  traffic for these contents is ridden with redundant requests \cite{itube1}.    
Thus, multicasting is a natural way to address these requests. 

A multicast queue with network coding is studied in \cite{Moghadam,Hou} with infinite library of files. The case of broadcast systems with one server transmitting to multiple users is studied in \cite{Cogil,su2000joint}. Both of these works study a slotted system. Some recent works \cite{Maddah-Ali2014} use coded caching to achieve multicast. This approach uses local information in the user caches to decode the coded transmission and provides improvement in throughput by increasing the effective number of files transferred per transmission. This throughput may get reduced in a practical scenario due to queueing delays at the basestation/server. \cite{Rezaei2016a} addresses these issues, analyses the queuing delays and compares it with an alternate coded scheme with LRU caches (CDLS) which provides improvement over the coded schemes in \cite{Maddah-Ali2014}. A more recent work in this direction, provides alternate multicast schemes and analyses queueing delays for such multicast systems \cite{Arxiv2018}. The authors show that a simple multicast scheme, can have significant gains over the schemes in \cite{Maddah-Ali2014}, \cite{Rezaei2016a} in high traffic regime. 

We further study the multicast scheme proposed in \cite{Arxiv2018} in this paper.  This multicast queue merges the requests for a given file from different users, arriving during the waiting time of the initial requests. The merged requests are then served simultaneously.  The gains achieved by this simple multicast scheme, however, are quickly lost in wireless channels due to fading. \cite{wcnc} addresses this issue and also provides several multicast queueing schemes to improve the average user delays. Also, it shows that these schemes combined with an optimal power control policy under average power constraint, can provide significant reduction in delays. 

The power control policy proposed in \cite{wcnc}, though provides improved delays, has following limitations:
\begin{itemize}
\item The algorithm to get the policy is not scalable with the number of users and the number of states of the channel gains.
\item The policy doesn't adapt to the changing system statistics, which in turn depends on the policy.
\end{itemize}           

\indent These systems are often conveniently modelled as a Markov Decision Process, but with large state and action spaces. Obtaining transition probabilities and the optimal policy for such large MDPs is not feasible. Reinforcement learning, particularly, Deep reinforcement learning comes as a natural tool to address such problems \cite{DBLP:journals/corr/abs-1810-06339}. Reinforcement learning has the added advantage that it can be used even when the transition probabilities are not available. However, large state/action space can still be an issue. Using function approximation via deep neural networks can provide significant gains  since the Q-values of different state-action pairs can be interpolated even if that state-action pair has never or rarely occurred in the past. Several, deep reinforcement learning techniques such as Deep Q-Network \cite{Mnih2015}, Trust Region Policy Optimization (TRPO) \cite{DBLP:journals/corr/SchulmanLMJA15}, Proximal Policy Gradient (PPO) \cite{DBLP:journals/corr/SchulmanWDRK17} etc. have been successfully applied to several large state-space dynamical systems such as Atari \cite{DBLP:journals/corr/MnihKSGAWR13}, AlphaGo \cite{Silver1140} etc. DQN is one of the first Deep-RL methods based on value iteration, usually employing $\epsilon$-greedy exploration to learn the optimal policy. TRPO and PPO are policy gradient based methods that employ stochastic gradient descent over policy space to obtain the optimal value function. Policy-Gradient methods often suffer from high variance in sample estimates and poor sample efficiency \cite{DBLP:journals/corr/abs-1810-06339}. Value iteration based deep RL methods, like DQN, have been theoretically shown to have competitive performance \cite{DBLP:journals/corr/abs-1901-00137}, specifically due to sample efficiency of experience replay \cite{Lin1992}.  
	
\indent In addition to the above mentioned trade-offs, a constrained stochastic optimization problem, as considered in this paper, further adds to the complexity of the problem. A modification of TRPO for constrained optimization is Constrained Policy Optimization \cite{DBLP:journals/corr/AchiamHTA17}. But, this too suffers from the high estimator variance issue. Work in \cite{RCPO} considers a multi-timescale approach for constrained DeepRL problems, as considered in this paper. However, \cite{RCPO} does not track the system statistics and hence cannot be applied in practical systems. Thus we propose a constrained optimization variant of DQN based on multi-timescale stochastic gradient descent \cite{borkar}. We have preferred DQN in this work, as the Target network and Replay memory used in the DQN reduce the estimator variance and finally achieve the global minimum empirical risk \cite{DBLP:journals/corr/abs-1901-00137}. 
	 
	The major contributions of this paper are:
\begin{itemize}
	\item Proposing two modifications to DQN, to accommodate constraints and system adaptations. We call this Adaptive Constrained DQN (AC-DQN).
	\item Unlike DQN, constrained DQN can be applied to the multicast systems with constraints, as in \cite{wcnc}, to learn the optimal power control policy, online. The constraints can be met by using a Lagrange multiplier. The appropriate Lagrange multiplier is also learnt via a two time scale stochastic gradient descent. The proposed method meets the average power constraint while achieving the global optima as achieved by the static policy proposed in \cite{wcnc}. 
	\item We demonstrate the scalability of our algorithms with system size (number. of users, arrival rate, complex fading).
	\item We show that AC-DQN can track the changes in the dynamics of the system, e.g., change of rate of arrival over the time of a day, and achieve optimal performance. 
\end{itemize}


Our algorithms work equally well when we replace DQN with its improvements such as DDQN \cite{ddqn}. In fact we have run our simulations with DDQN variant of AC-DQN and have achieved similar performance. Next, we describe some more related works to this paper:

\textbf{Power control in Multicast Systems:} Power control in multicast systems has been studied in \cite{Goldsmith, Proakis, Kim}. In \cite{Goldsmith}, optimal power allocation is made to achieve the ergodic capacity (defined as the boundary of the region of all achievable rates in a fading channel with arbitrarily small error probability) while maintaining minimum rate requirements at users and average power constraints.   Authors use water-filling to achieve the optimal policy. In \cite{Proakis}, the authors minimize a utility function via linear programming, under SINR constraints at the users and transmit power constraints at the transmitter. Both \cite{Goldsmith,Proakis} derive an optimal power control policy for delivery to all the users, whereas this paper considers delivery to a random subset of users. In \cite{Kim}, each packet has a deadline and packets not received by the end of the slot are discarded. The authors use dynamic programming to obtain the optimal policies.

\textbf{Deep Reinforcement Learning (DeepRL) in Wireless Multicast systems:} The ability of DeepRL to handle large state-space dynamic systems is being exploited in various multicast wireless systems/networks. In \cite{Hao}, authors look at resource allocation problem in unicast and broadcast. The DeepRL agent learns and selects power and frequency for each channel to improve rate, under some latency constraints. Authors, like in our work, introduce constraints via Lagrange multiplier.   However, the agent doesn't learn the Lagrange multiplier.   Thus, the agent also does not adapt if the system dynamics changes as the  Lagrange constant in the reward is fixed for a given dynamics,   and the learning rate decays with time. Another work, \cite{Nasir}, applies unconstrained deep reinforcement learning to multiple transmitters for a proportionally fair scheduling policy by adjusting individual transmit powers. Some studies \cite{Dai} have applied DeepRL to control power for anti-jamming systems.  

\indent Rest of the paper is organised as follows. Section \ref{sec:system_model} explains the system model and motivates the problem. Section \ref{sec:DRL_power_cont } presents the proposed DeepRL algorithm AC-DQN. Section \ref{sec:simulation} demonstrates our algorithms via simulations and Section \ref{sec:conclusion} concludes the paper. \section{System Model} 
\label{sec:system_model}
\begin{figure}[h!]
\centering
\includegraphics[height=5cm,width = 7cm]
{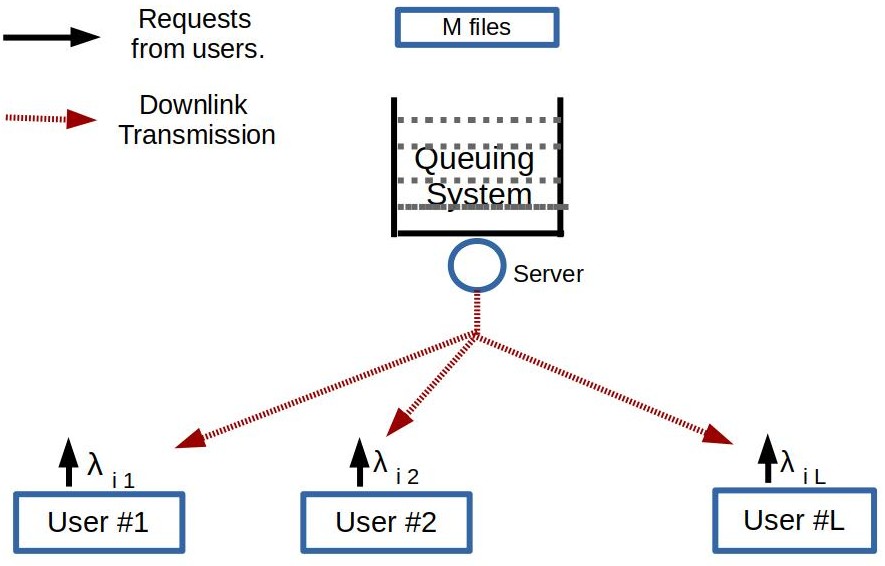}
\caption{System model}
\label{fig:sys_model}
\end{figure}

We consider a system with one server transmitting files from a fixed finite library to a set of users (Figure \ref{fig:sys_model}). We denote the set of users by $\mathcal{L} = \{1,2,\cdots, L\}$ and the set of files by $\mathcal{M} = \{1,2,\cdots, M\}$. We assume that $M >> L$. The request process for file $i$ from user $j$ is a Poisson process of rate $\lambda_{ij}$ which is independent of the request processes of other files from user $j$ and also from other users. The total arrival rate is $\lambda=\sum_{i,j}\lambda_{ij}$. The requests of a file from each user are queued at the server till the user successfully receives the file. All the files are of length $F$ bits. The server transmits at a fixed rate, $R$ bits/sec. Thus, the transmission time for each file is $T = F/R$. 
 
\indent The channels between the server and the users experience time varying fading. The channel gain of each user is assumed to be constant during transmission of a file. The channel gain for the $j^{th}$ user at the $t^{th}$ transmission, is represented by $H_j(t)$. Each $H_j(t)$ takes values in a finite set and form an independent identically distributed (i.i.d) sequence in time, as in \cite{fsmc}. The  channel gains of different users are independent of each other and may have different distributions. Let ${H}=\{H_1, \cdots, H_L\}$.
 
 More details of the system are described in the following subsections as follows. Section \ref{sec:m_queue} describes the basic Multicast queue proposed in \cite{Arxiv2018}. The scheduling scheme to mitigate the effects of fading studied in \cite{wcnc} are also presented. In Sections \ref{sec:avg_pow_cons} and \ref{sec:MADS} we summarise  the results from \cite{wcnc} which show that using power control can further improve the performance and the algorithm used to obtain the optimal power policy.  We will see that this algorithm is not scalable. Then in Section \ref{sec:MDPF}  we provide the MDP of the power control problem. In Section \ref{sec:DRL_power_cont } we will present the scalable  DeepRL  solution for this formuation. 
 
\subsection{Multicast Queue}
\label{sec:m_queue} 
\indent For scheduling of transmission at the server, we consider the \textit{multicast queue} studied in \cite{wcnc}. In this system, the requests for different files from different users are queued in a single queue, called the multicast queue. In this queue, the requests for file $i$ from all users are merged and considered as a single request. The requested file and the users requesting it, is denoted by $(i,\mathbb{L}_i$). A new request for file $i$, from user $j$ is merged with the corresponding entry $\mathbb{L}_i$, if it exists. Else, it is appended to the tail of the queue. Service/transmission of file $i$, serves all the users in $\mathbb{L}_i$, possibly with errors due to channel fading.\\
\indent The random subset of users served by the multicast queue at the $t^{th}$ transmission, is denoted by the random binary vector, ${V}(t)=\{V_1(t), \cdots, V_L(t)\}$, where $V_j(t)=1$ implies that the user $j$ has requested the file being transmitted; otherwise, $V_j(t) = 0$. From [Theorem 1, \cite{Arxiv2018}], ${V}(t)$ has a unique stationary distribution. 

\indent It was shown in \cite{Arxiv2018} that the above multicast queue performs much better than the multicast queues proposed in literature before. The main difference compared to previous multicast schemes is that in this scheme, all requests of all the users for a given file are merged together over time. One direct consequence of this is that the queue length at the base station does not exceed $M$. Thus the delay is bounded for all traffic rates. In fact the mean delays are often better than the coded caching schemes proposed in the literature, as well, for most of the traffic conditions. However, in a fading scenario, where the different users have independent fading, the performance of this scheme can significantly deteriorate because of multiple retransmissions required to successfully transmit to all the users needed. Thus, in \cite{wcnc}, multiple retransmission schemes were proposed and compared to recover the performance of the system. The following scheme was among the best. It not only (almost) minimizes the overall mean delays of the system, it also is fair to different users in the sense, that the users with good channel gains do not suffer due to users with bad channel gains.
 
\textbf{Single queue with loop-back (1-LB):} The Multicast queue is serviced from head to tail. When a file is transmitted, some of the users will receive the file successfully and some users may receive the file with errors. In the case of unsuccessful reception by some users, the file is retransmitted. A maximum of $N$ $(1 \leq N \leq \infty)$ transmission attempts are made. If there are some users who did not receive the file within $N$ transmission attempts, the request (tuple $(i,\mathbb{L}_i)$ with $\mathbb{L}_i$,   now modified to contain only   the set of users who have not received the file $i$ successfully) is fed back to the queue. If there is another pending request in the queue for the same file (a request for the file which came during the transmission of the current transmission), it is merged with the existing request. Otherwise, a new request for the same file with unsuccessful users is inserted at the tail of the queue.

\indent It was further shown in \cite{wcnc} that  choosing the transmit power based on the channel gains, can further improve the system performance.
\subsection{Average Power Constraint}
\label{sec:avg_pow_cons}
\indent Depending on the value of ${H}(t)$ and ${V}(t)$ at time $t$, the server chooses transmit power $P_t$, based on a power control policy $P_t=\pi({H}(t),{V}(t))$. Choosing a good power control policy is the topic of this paper. 

	The state, $S_t$  of the system at time $t$ is $({H}(t), {V}(t))$. Let $\ P_{S_t}$ be the power chosen by a policy for state $S_t$ and $R(S_t, P_{S_t})$ be the number of successful transmissions for the selected power $P_{S_t}$, during the $t^{th}$ service.\\
\indent  For a fixed transmission rate $C$ and for a given  channel gain $H(t)$ of users, the transmit power requirement $P_{req}$ (from Shannon's Formula) for user $j$ is  (assuming file length is long enough) 
\begin{equation}\label{eq:pow_req}
	P_{req}(j,S_t)=\frac{N_g}{H_j^2(t)}(2^{C/B}-1),
\end{equation}	      
where, $B$ is the bandwidth and $N_g$ is the Gaussian noise power at receiver $j$. Thus the reward for the chosen power control policy, during $t^{th}$ transmission is given by,
\begin{equation}\label{eq:reward}
	R({S_t},P_{S_t})=\sum_{j=1}^L{V_{j,S_t}\ 1_{\{P_{S_t}>P_{req}(j,S_t)\}}}(t),
\end{equation}
where $V_{j,S_t} = 1$ if the user $j$ has requested the file in service and $V_{j,S_t} = 0$ otherwise. We now describe the Mesh Adaptive Direct Search (MADS) power control policy.
 
\subsection{MADS Power control policy}
\label{sec:MADS}
  The power control policy in \cite{wcnc} is derived from the following optimization problem,
\begin{equation}
\underset{\{P_1,\cdots,P_K\}}{\max}{\sum_{k=1}^K{q_k R_k}}
\end{equation}
\begin{equation}
s.t.\ {\sum_{k=1}^K{q_k P_k}}\leq \overline{P}\ \text{and}\ P_k\geq 0, k=1,\cdots,K,
\end{equation}
where, $\overline{P}$ is the average power constraint, $K$ is the total number of states, $P_k$ is the power chosen by the policy in state $k$, $q_k$ is the stationary distribution of state $k \in \{1,\cdots,K\}$ and $R_k$ is the reward for state $k$, as defined in (\ref{eq:reward}) with $S_t=k$. This is a non-convex optimization problem. Mesh Adaptive Direct Search (MADS) \cite{MADS} is used in \cite{wcnc} to solve this constrained optimization problem and obtain the power control policy. Though MADS achieves global optimum, it is not scalable as its computational complexity is very high.

	The state space and action space of this problem can be very high even for a moderate number of users and channel gains, e.g., a system with L users and G channel gain states, has $\mathcal{O}(2^L G^L)$ states. Therefore, in this paper we propose a deep reinforcement learning framework. This not only provides optimal solution for a reasonably large system but does so without knowing the arrival rates and channel gain statistics. In addition, we will be able to provide an optimal solution even when the arrival and channel gain statistics change with time. 

\subsection{MDP Formulation}
\label{sec:MDPF}
The above system can be formulated into a finite state, action Markov Decision Process denoted by tuple ($\mathbb{S}, \mathbb{A}, r, \textbf{P}, \gamma$): (state space, action space, reward, transition probability, discount factor), where, transition probability $\textbf{P}(S_{t+1}|S_0,P_0, ..., S_t, P_t)=\textbf{P}(S_{t+1}|S_t, P_t)$, policy $\pi$  chooses power $P_t \sim \pi (.|S_t)$ in state $S_t$ and the instantaneous reward $r_t=R({S_t},P_{t})$.\\  
The action-value function \cite{Puterman:1994:MDP:528623} for this discounted MDP for policy $\pi$ is
\begin{equation}
\begin{split}
Q^{\pi}&(s,a)=\mathbb{E}[\sum_{t=0}^{\infty}{\gamma^t r_t}|S_0=s,P_0=a].
\end{split}
\end{equation}
The optimal $Q\text{-function}$, $Q^*$ is given by $Q^*(s,a)=\underset{\pi}{\max}\ Q^{\pi}(s,a)$ and satisfies the optimality relation,

\begin{equation}\label{eq:bellman}
Q^*(s,a)=r(s,a)+\underset{a'}{\max}\ {\gamma}\mathbb{E}[Q^*(s',a')],
\end{equation}  
where, $s'$ is sampled with distribution $\textbf{P}(.|s,a)$. If we know the optimal Q-function $(Q^{*})$, we can compute the optimal policy via $\pi(s)=\underset{a'}{\arg\max}\ Q^{*}(s,a)$. We know the transition matrix of this system and hence can compute the $Q$-function. But the state space is very large even for a small number of users, rendering the computations infeasible.  Thus, we use a parametric function approximation of the Q function via Deep neural networks and use DeepRL algorithms to get the optimal $Q^{*}$.  

Further, to introduce the constraint in the MDP formulation, we look at the policies achieving 
\begin{equation}
Q^*(s,a)=\underset{\pi:C_P\leq \overline{P}}{\max}\ Q^{\pi}(s,a)
\end{equation}
where
\begin{equation}\label{eq:long_avg}
C_P=\mathbb{E}[\underset{T\rightarrow\infty}{\lim} \frac{\sum_{t=0}^T P_t}{T}]
\end{equation} 
is the long term average power. We use the Lagrange method for constrained MDPs \cite{Altman} to achieve the optimal policy. In this method, the instantaneous reward is modified as 
\begin{equation}\label{eq:lagrange_cmdp}
	r_t=R(S_t,P_t)-\beta P_t,
\end{equation}    
where, $\beta$ is the Lagrange constant achieving optimal $Q^{*}$ while maintaining, $C_P \leq \overline{P}$. Choosing $\beta$ wrongly will provide the optimal policy with average power constraint different from $\overline{P}$.  
\CE
\section{Deep Reinforcement Learning based Power Control Policy}
\label{sec:DRL_power_cont }
\indent In this section, we describe the Deep-Q-Network (DQN) \cite{Mnih2015} based power control. First we describe the DQN algorithm. We then propose a variant of DQN for constrained problems, where in, we use a Lagrange multiplier to take care of the average power constraint. We use multi-timescale stochastic gradient descent approach to learn the Lagrange multiplier, to obtain the right average power constraint. Finally, we change the learning step size from decreasing to a constant so that the optimal power control can track the time varying system statistics. 
\subsection{Deep Q Networks}
DQN is a popular Deep Reinforcement learning algorithm to handle large state-space MDPs with unknown/complex dynamics, $\textbf{P}(S_{t+1}|S_t, P_t)$. The DQN is a Value Iteration based method, where the action-value function is approximated by a Neural Network. Though there are several follow up works providing improvements over this algorithm \cite{ddpg, ddqn}, we use this algorithm owing to its simplicity. We will show that DQN itself is able to provide us the optimal solution and tracking. These improvements may further improve the performance in terms of sample efficiency, estimator variance etc. The DQN algorithm is given in Algorithm \ref{algo:dqn}. Earlier attempts in combining nonlinear function approximators such as neural networks and RL were unsuccessful due to instabilities caused by 1) correlated training samples, 2) drastic change in policy with small change in function approximation, and 3) correlation between the training function and approximated function \cite{DBLP:journals/corr/abs-1810-06339}. Success of DQN is attributed to addressing these issues with two key ingredients of the algorithm: \textbf{Experience Replay Memory} $\mathbb{M}$ and \textbf{Target Network}, $Q_{\theta^{*}}$. The replay memory stores the transitions of an MDP, specifically the tuple, $(S_t, P_t, r_t, S_{t+1})$. The algorithm then samples, uniformly, a random minibatch of transitions from the memory. This removes correlation  between the data and smoothens the data distribution change with iteration.  The target network and randomly sampled mini-batch from the memory $\mathbb{M}$, form the training set for training the $Q_{\theta}$ Network, at every epoch. This random sampling provides $i.i.d$ samples for performing stochastic gradient descent with {loss} 
\begin{equation}
L^{\pi_{\theta}}_{Q} = \frac{1}{n}\sum_{j=1}^n(Y_j-Q_{\theta}(S_j, A_j))^2
\end{equation}
 where, $Y_i=r_i+\gamma\ \underset{a'}{\max}Q_{\theta^{*}}(S_i,a'))$. The iterates $\{\theta_t\}$ are given by: 
\begin{equation}
\theta_{t+1}\leftarrow\theta_t - \eta_1(t) \nabla_{\theta} L^{\pi_{\theta}}_{Q},
\end{equation}
where $\eta_1(t)$ satisfies:
\begin{equation}
\label{eq:step_size_dqn}
\sum_{t=0}^{\infty}\eta_1(t)=\infty,\ \sum_{t=0}^{\infty}\eta_1^2(t)<\infty,\ \eta_1(t)\geq 0. 
\end{equation}
The weights of the target network $Q^{*}$ are held constant for $T_{target}$ epochs, thereby controlling any drastic change in policy and reducing correlation between $Q$ and $Q^{*}$. This can be seen as a Risk Minimization problem in nonparametric-regression with {regression function} $Q_{\theta^{*}}$ and {risk} $L^{\pi_{\theta_t}}_{Q}$. Readers are referred to \cite{DBLP:journals/corr/abs-1901-00137} for elaborate analysis of DQN.  Theorem 4.4 in \cite{DBLP:journals/corr/abs-1901-00137} provides a proof of convergence and the rate of convergence using non-parametric regression bounds, when sparse ReLU networks are used, under certain smoothness assumptions on the reward function and the dynamics.

\begin{algorithm}\label{algo:dqn}
\SetAlgoLined
\caption{Deep-Q-Network}
\KwIn{MDP-$(\mathbb{S},\mathbb{A}, r, \textbf{P}, \gamma)$, Replay Memory $\mathbb{M}$, Minibatch size:$n$ $T$, $T_{target}$, Initialize weights $\theta, \theta^{*}$ of $Q_{\theta}$ and $Q_{\theta^{*}}$, $\epsilon(t)\rightarrow 0$: Exploration Parameter, $\eta_1(t)$: Learning rate satisfying (\ref{eq:step_size_dqn})}
\For{$t=1$ \KwTo $T$}{
	Observe state $S_t$,	Apply action $A_t=\pi_t(S_t)=\underset{a'}{\arg\max\ }{Q_{\theta}(S_t,a')}$, $\epsilon$-greedily\\
	Observe: $r_t,S_{t+1}$ \\ Store: $(S_t, A_t, r_t, S_{t+1})$ in $\mathbb{M}$\\
	\textbf{Sample:} Minibatch $n$ from $\mathbb{M}$\\
	\For{$i=1$ \KwTo $n$}{
	$Y_i=r_i+\gamma\ \underset{a'}{\max}\ Q_{\theta^{*}}(S_{i+1},a')$
	}
$\theta\leftarrow \theta - \eta_1 \nabla_{\theta} L^{\pi_{\theta}}_{Q}$\\
	at every ${t=m T_{target}}, {m\in \mathbb{N^{+}}}$: update $\theta^{*}\leftarrow{\theta}$
}
$\pi^{*}\leftarrow\pi_T$, $\theta^{*}\leftarrow{\theta}$\\
\KwOut{$Q_{\theta^{*}}$: Optimal $Q$-Function, $\pi$: Optimal Policy}
\end{algorithm}  

  
\subsection{Adaptive Constrained DQN (AC-DQN)}
The DQN algorithm is meant for unconstrained optimization. Since our problem has an average power constraint of $\overline{P}$, we consider the instantaneous reward in (\ref{eq:lagrange_cmdp}), with a Lagrange multiplier $\beta$. The long term constraint depends on the Lagrange multiplier and can be quite sensitive to it. Thus, we design our algorithm, AC-DQN, to learn the appropriate $\beta$. We will see later, that this will enable us to further modify our algorithm to track the changing statistics of the  channel gains and arrival statistics. The AC-DQN algorithm is given in Algorithm \ref{algo:AC-DQN}. Here, we use multi-timescale SGD as in \cite{borkar}. In this approach, in addition to the SGD on $Q_{\theta}$, using minibatch, we use a stochastic gradient descent on the Lagrange constant, $\beta$ as 
\begin{equation}
\beta_{t+1} \leftarrow \beta_{t} + \eta_2(t) \nabla_{\beta} L_P^{{\pi}_{\theta}}, 
\end{equation}
where $\nabla_{\beta} L_P^{{\pi}_{\theta}} = C_P(S_t)-\overline{P}$. Since the expectation in (\ref{eq:long_avg}) is not available to us, we take $C_P(S_t)={\sum_{i=t-T_W}^t P_i(S_i)}/{T_W}$, where $T_W$ is the finite horizon window.
Additionally $\eta_1$ and $\eta_2$ are required to follow \cite{borkar}:
\begin{equation}
\begin{split}
\label{eq:two_timeline_learning}
\sum_{i=1}^{\infty}\eta_1(i)=\sum_{i=1}^{\infty}& \eta_2(i)=\infty,\\ \sum_{i=1}^{\infty}\eta_1^2(i)+\eta_2^2(i)&<{\infty},\ \frac{\eta_2(i)}{\eta_1(i)}\rightarrow 0.
\end{split}
\end{equation}

\begin{algorithm}\label{algo:AC-DQN}
\SetAlgoLined
\caption{Adaptive Power Control DQN (AC-DQN) Algorithm}
\KwIn{MDP-$(\mathbb{S},\mathbb{A}, r, \textbf{P}, \gamma)$, $r$ as in (\ref{eq:lagrange_cmdp}), Replay Memory $\mathbb{M}$, Minibatch size: $n$ $T$, $T_{target}$, Initialize weights $\theta, \theta^{*}$ of $Q_{\theta}$ and $Q_{\theta^{*}}$, $\epsilon(t)\rightarrow 0$: Exploration Parameter, $\beta$: Lagrange Constant, $\eta_1(t)$: Value learning rate, $\eta_2(t)$: Lagrange learning rate satisfying (\ref{eq:two_timeline_learning}), Initialize $T_W$}
\For{$t=1$ \KwTo $T$}{
	Observe state $S_t$,	Apply action $A_t=\pi_t(S_t)=\underset{a'}{\arg\max\ }{Q_{\theta}(S_t,a')}$, $\epsilon$-greedily\\
	Observe: $r_t,S_{t+1}$ \\ Store: $(S_t, A_t, r_t, C_P(S_t), S_{t+1})$ in $\mathbb{M}$\\
	\textbf{Sample:} Minibatch $n$ from $\mathbb{M}$\\
	\For{$i=1$ \KwTo $n$}{
	$Y_i=r_i+\gamma\ \underset{a'}{\max}Q_{\theta^{*}}(S_{i+1},a')$
	}
$/*${Perform two time-scale stochastic gradient descent as follows:}$*/$\\	
$\theta\leftarrow \theta - \eta_1 \nabla_{\theta} L^{\pi_{\theta}}_{Q}$\\
$\beta \leftarrow \beta + \eta_2 \nabla_{\beta}L^{\pi_{\theta}}_{P}$ \\
	at every ${t=m T_{target}}, {m\in \mathbb{N^{+}}}$: update $\theta^{*}\leftarrow{\theta}$
}
$\pi^{*}\leftarrow\pi_T$, $\theta^{*}\leftarrow{\theta}$\\
\KwOut{$Q_{\theta^{*}}$: Optimal $Q$-Function, $\pi$: Optimal Policy}
\end{algorithm}  

\textbf{Tracking with AC-DQN:} Tracking of system statistics is essential, to achieve optimal power control in a non-stationary system. In multi-time scale stochastic gradient descent, such as AC-DQN, step sizes $\eta_1(t)$ and $\eta_2(t)$ can be fixed to enable tracking. If $\eta_2<<\eta_1$, then the Lagrange multiplier changes much more slowly than the $Q$-function. Then the two timescale theory (see, e.g., \cite{borkar}), will allow the Lagrange multiplier to adapt slowly to the changing system statistics but at the same time provide average power control. The solution will reach in a neighbourhood of the optimal point.

Although the convergence of this modified algorithm is not proved yet (even for the unconstrained DQN, convergence has been proved only recently  in \cite{DBLP:journals/corr/abs-1901-00137}), our simulations will show that the resulting algorithm tracks the optimal solution in the time varying scenario.

The time varying scenario in our setup results due to change in the request arrival statistics from the users and changing channel gain statistics due to motion of the users.
\section{Simulation Results and Discussion}
\label{sec:simulation}
\indent In this section, we demonstrate the Deep Learning methods for power control proposed in this paper. We compare performances of AC-DQN and MADS Power control policies. Though MADS provides optimal solutions for small system sizes, it is not scalable. We show that the Deep Learning algorithm, AC-DQN, indeed achieves the global optimum obtained by MADS algorithm, while being scalable with the system size (number of users). We further demonstrate that AC-DQN algorithm tracks the changing system dynamics and obtains the optimal policy, adaptively. We use Keras libraries \cite{keras} in Python for implementation of our algorithms and our system is implemented in MATLAB.

We consider two systems, one with 4 users and compare AC-DQN and MADS; the other with 20 users, showing performance of AC-DQN. MADS is not able to provide a solution for the second system since the space and time complexity of MADS increase exponentially with the number of users. In all the examples, we split the users in two equal sized groups, one group has good channel statistics and the other bad channel statistics. In both the systems, we compare all the algorithms with a constant power control policy, where the transmit power $P_t$ is fixed to $P_t=\overline{P}$, to indicate the gain due to power control. The system and algorithm parameters, used for the simulations are as follows:  

\subsubsection{\textbf{Small User Case}}\label{sec:system_4} Number of users, $L=4$, Catalog Size $M= 100$, File Size $F=10MB$, Transmission rate $C=10MB/s$, Bandwidth $B=10MHz$, Channel Gains, $\sim$ Uniform([0.1 0.2 0.3]) for two users with bad channel statistics and $\sim$ Uniform([0.7 0.8 0.9]) for two users with good channel statistics, File Popularity : Uniform, (Zipf exponent = 0). Average Power Constraint  $\overline{P}=7$, Simulation time= $10^5$ mutlicast transmissions.

\subsubsection{\textbf{Large User Case}}\label{sec:system_20} System Parameters: Power Transmit Levels = 20 (1 to 50), $L=20$, $M = 100$, $F=10MB$, $C=10MB/s$, Channel Gains : Exponentially distributed. ($\sim exp(0.1)$ for bad channel, $\sim exp(1.0)$ for good channel), $R=10MB/s$, $B=10MHz$, $\overline{P}=7$, File Popularity : Zipf distribution with (Zipf exponent = 1). Simulation time: $10^5$ mutlicast transmissions. In both the cases we set the noise power as $N_g=1$.

\subsubsection{\textbf{Hyperparameters}}
\begin{figure}[h]
\centering
\includegraphics[trim={1cm 7.5cm 1cm 8cm},clip,height=5.5cm,width=8.5cm]{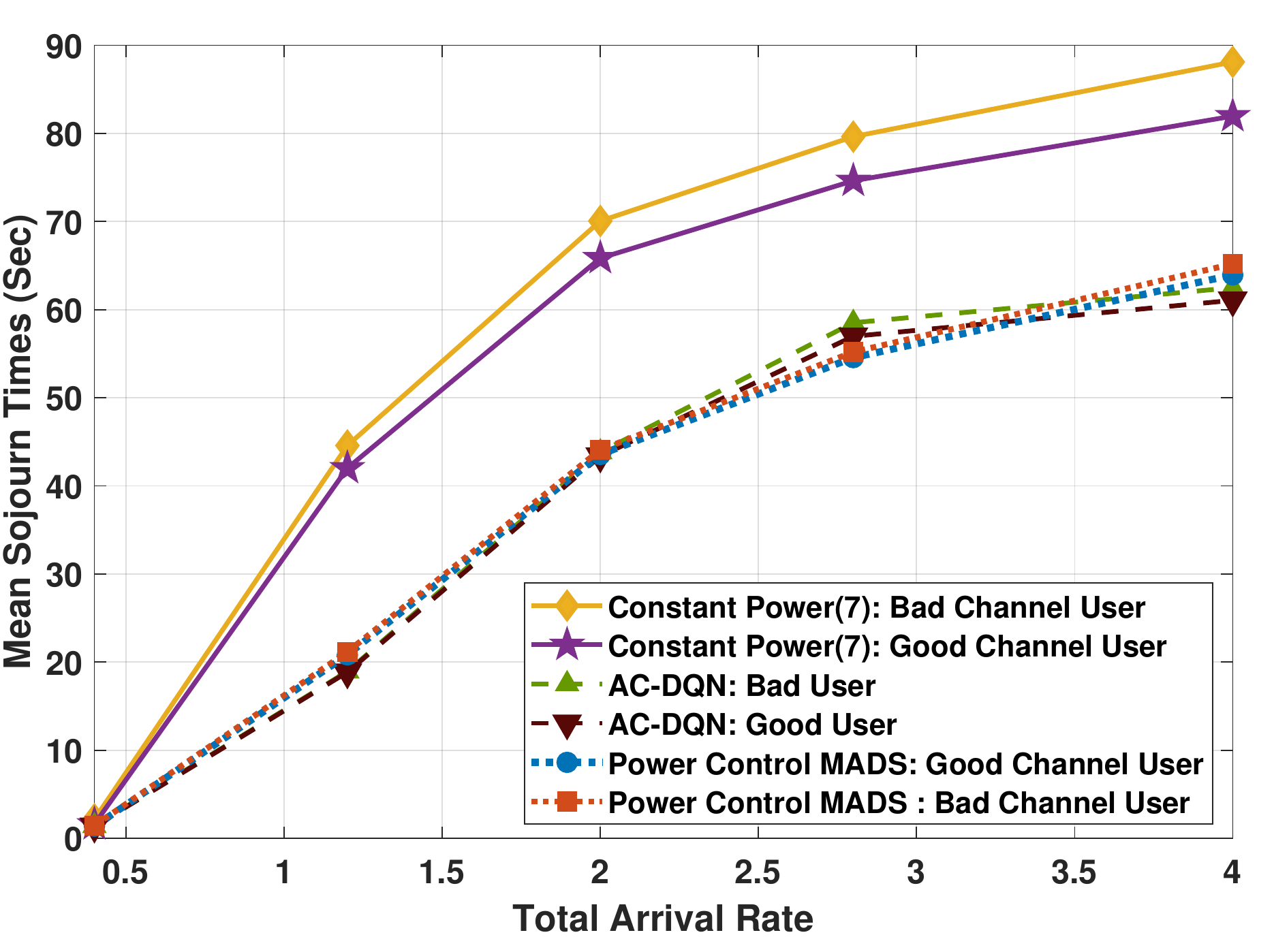}
\caption{Sojourn Times of MADS, PCD and AC-DQN vs Arrival Rate. $L=4$, $\overline{P}=7$, Uniform Popularity, Uniform fading.}
\label{fig:mads_comp_del}
\end{figure}

\begin{figure}[!h]
\centering
	\includegraphics[trim={1cm 13.3cm 1.2cm 8.5cm},clip,height=3.5cm,width=8.5cm]{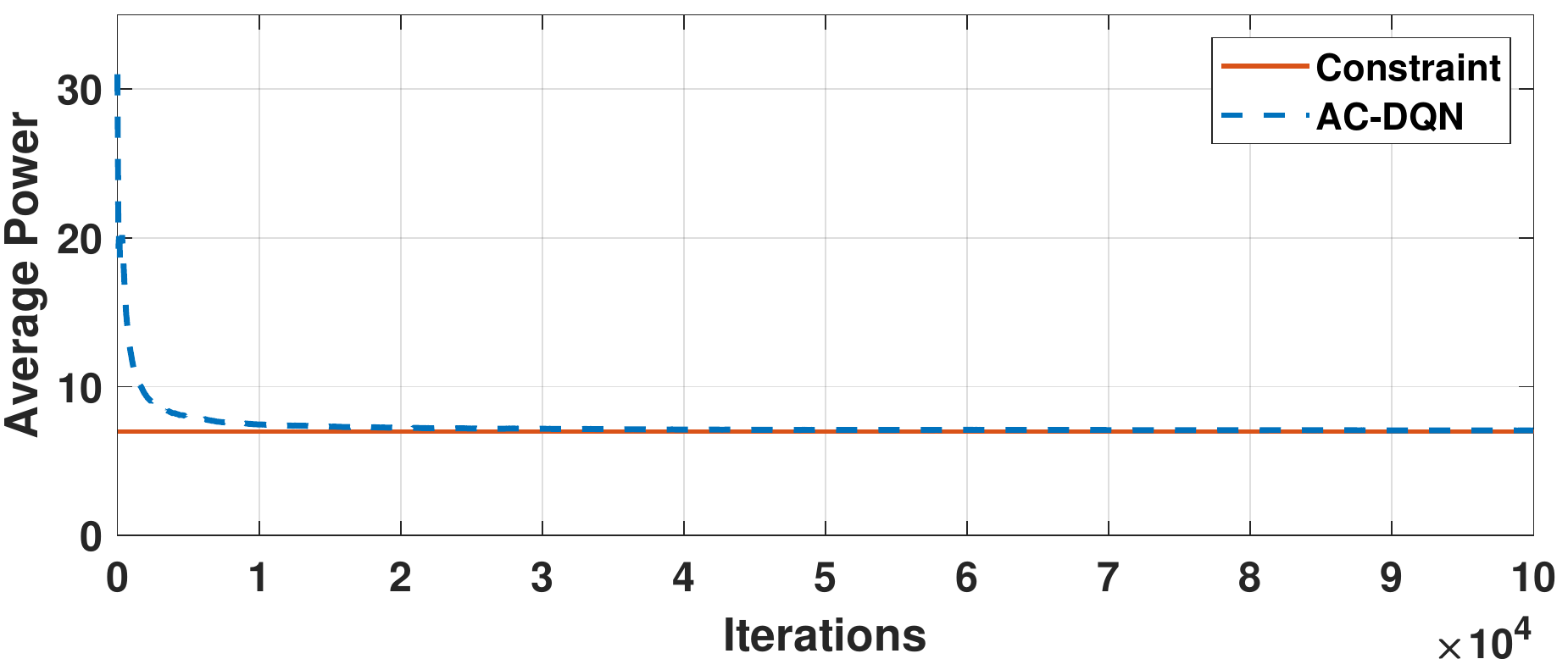}
\caption{Convergence of Average power with Iteration for AC-DQN, $L=4,\overline{P}=7$.}	
\label{fig:AC-DQN_power_4}
\end{figure}

We consider fully connected neural networks  with two hidden layers for all the function approximations considered in the algorithms.   Input layer nodes are assumed to be $2L$ and the output layer nodes is equal to 20, the number of transmit power levels. Each output represents the Q value for a particular action. The action space is restricted to be finite, as DQN converges only with finite action spaces.  We use two hidden layers for the neural network, with 128 and 64 nodes, and ReLU activation function is chosen, respectively. The other parameters are as follows: Replay memory size $\vert\mathbb{M}\vert=30000$,
$\gamma= 0.9$, $\epsilon_0=1.0$, $\epsilon_{decay}= 0.98$, $\epsilon_t = \epsilon_0 (0.98)^t$, $\eta_1= 0.001$, $\eta_1^{decay}=0.00001$, $\eta_2=.0001$, $\eta_2^{decay}=0.00001$, Mini-batch Size $(n)=64$, $T_{target}=100$, and $T_W=200$. \\
\textbf{Achieving Global Optima (AC-DQN vs MADS):}
  We use the system setting of small user case, specified above.  We run the system for the average power constraint ${\overline{P}=7}$, with exponential arrivals of rate 0.4 to 4.0. Figure \ref{fig:mads_comp_del} shows a comparison of sojourn times of Constant Power Policy, $P_t=\overline{P}$, MADS and AC-DQN. Further, Figure \ref{fig:AC-DQN_power_4} shows convergence of average power to $\overline{P}$ for AC-DQN. We see from Figure \ref{fig:AC-DQN_power_4} that AC-DQN achieves the global optimum achieved by MADS, while maintaining the average power constraint.

\begin{figure}[h!]
\includegraphics[trim={.7cm 7.6cm 1.2cm 8.1cm},clip,height=5.5cm,width=8.5cm]{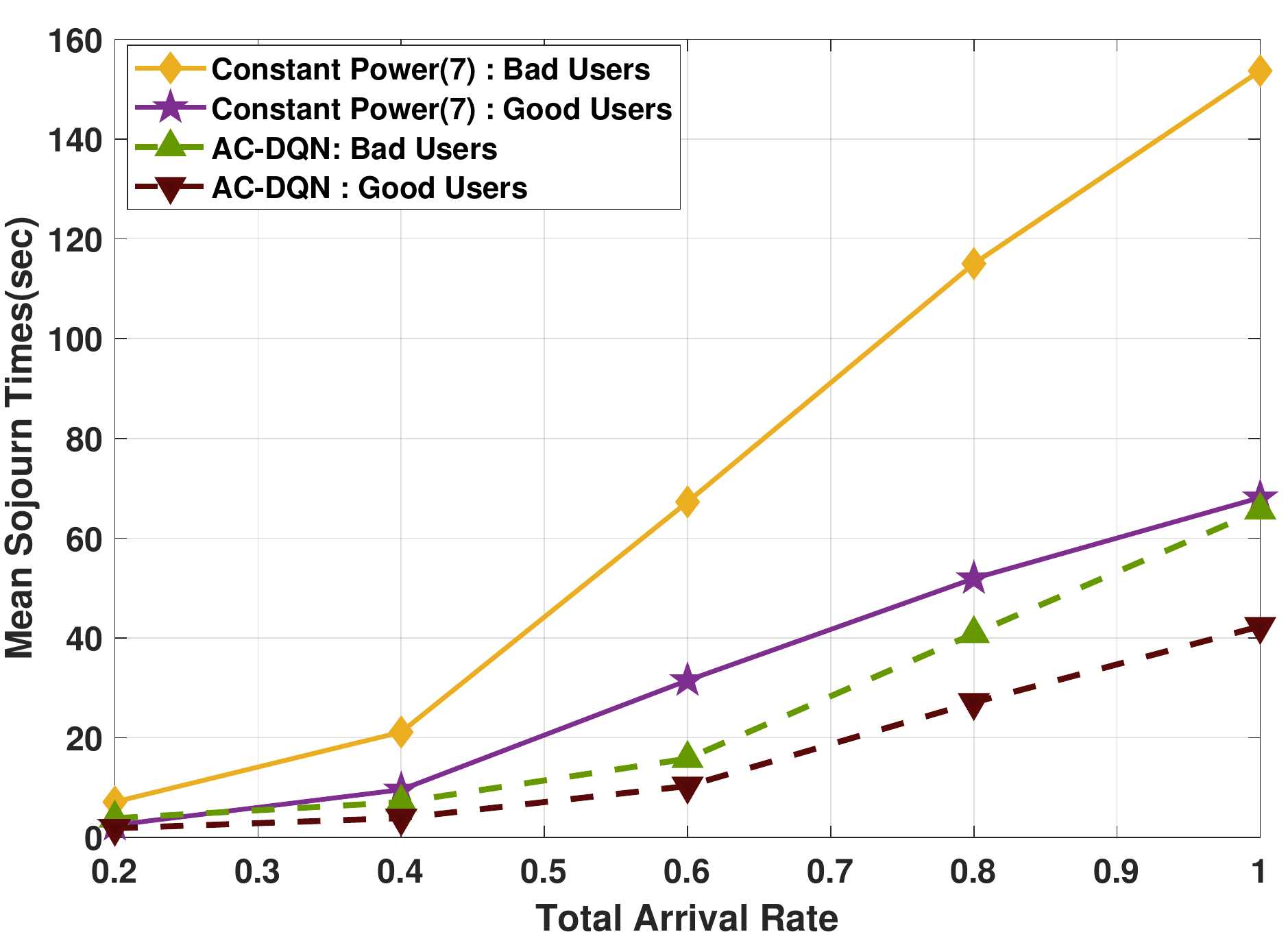}
\caption{Sojourn times for Constant Power and AC-DQN vs Arrival Rate. $L=20$, $\overline{P}=7$, Zipf(1) Popularity, Rayleigh fading.}
\label{fig:AC-DQN_comp_del}
\end{figure}

\begin{figure}[h!]
\centering
	\includegraphics[trim={1.1cm 13cm 1.4cm 8.5cm},clip,height=3.5cm,width=8.5cm]{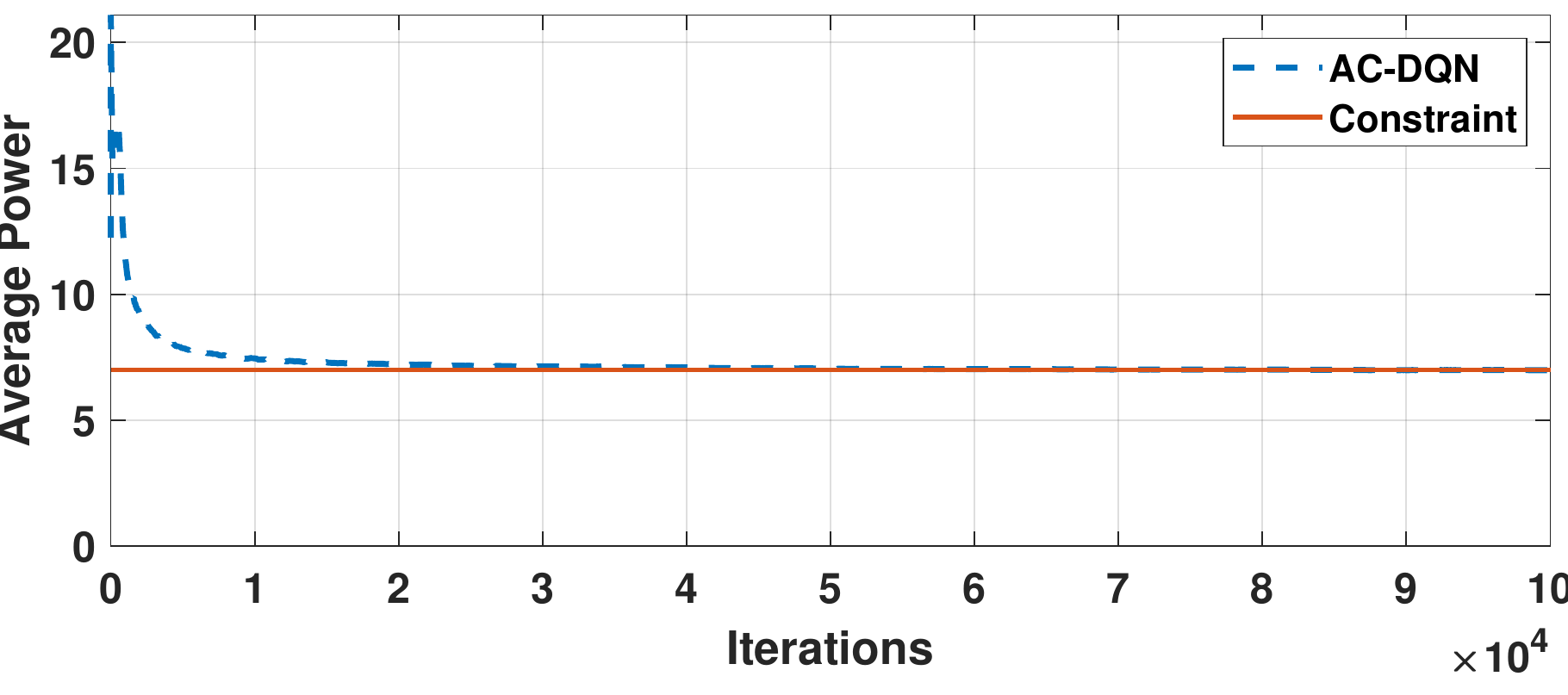}	
\caption{Convergence of Average Power, AC-DQN, $L=20, \overline{P}=7.$}
\label{fig:APCD_power_20}
\end{figure}

\begin{figure}[h!]
\centering
	\includegraphics[trim={.6cm 13.3cm 1.4cm 8.5cm},clip,height=3.5cm,width=8.5cm]{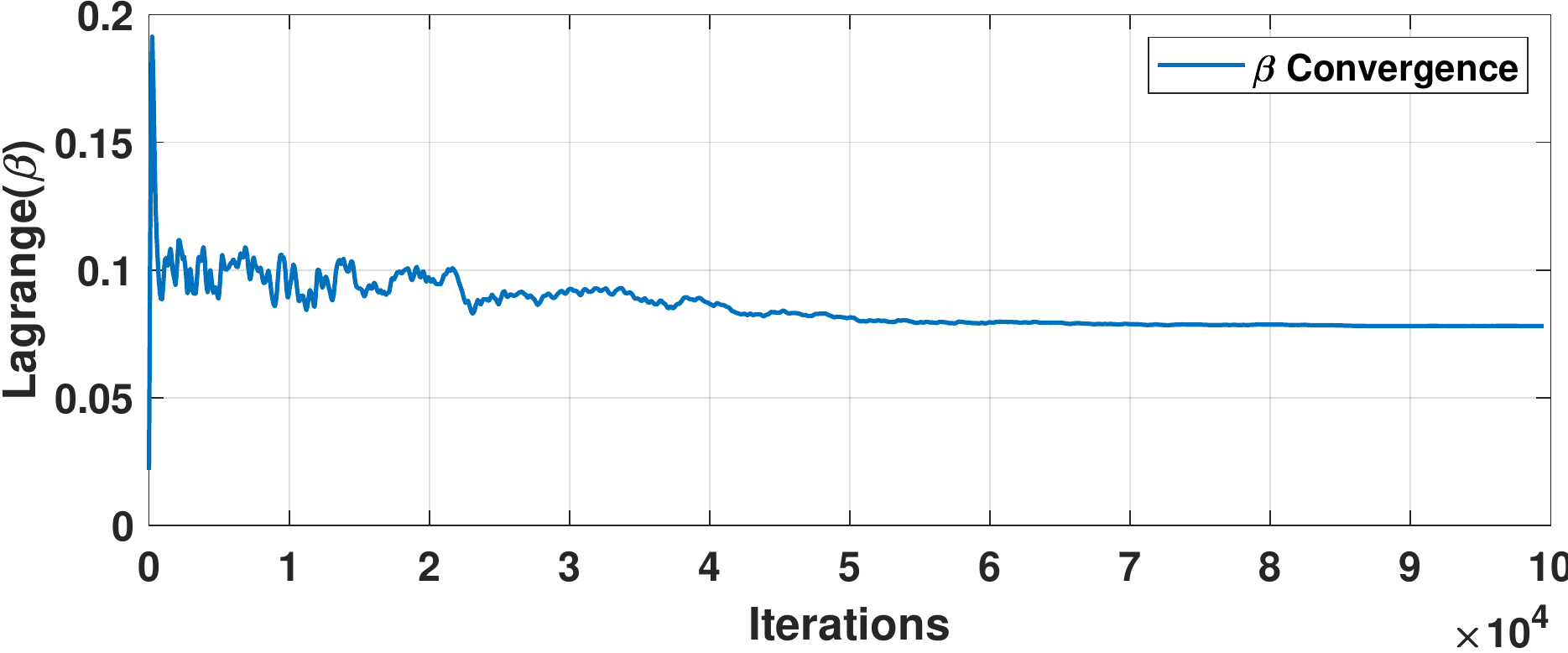}	
\caption{Convergence of Lagrange multiplier $L=20, \overline{P}=7.$}
\label{fig:Lagrange_20}
\end{figure}

\textbf{AC-DQN performance in a Scaled Network:}
To show the scalability of AC-DQN, we simulate  the relatively complex system mentioned in large user case, above. We run the simulation for $\overline{P}=7$. We see in Figure \ref{fig:AC-DQN_comp_del} that the AC-DQN gives, drastic improvement (around ~50 percent) over constant power case. AC-DQN achieves this while maintaining the average power, by learning the Lagrange constant as seen in Figure \ref{fig:Lagrange_20}. Figure \ref{fig:APCD_power_20} shows the convergence of average power of AC-DQN to the average power constraint, $\overline{P}$, for arrival rate of 1.0 requests per sec in the same simulation run.

\begin{figure}[h!]
\centering
\includegraphics[trim={.7cm 7.7cm 1.6cm 8cm},clip,height=5.5cm,width=8.5cm]{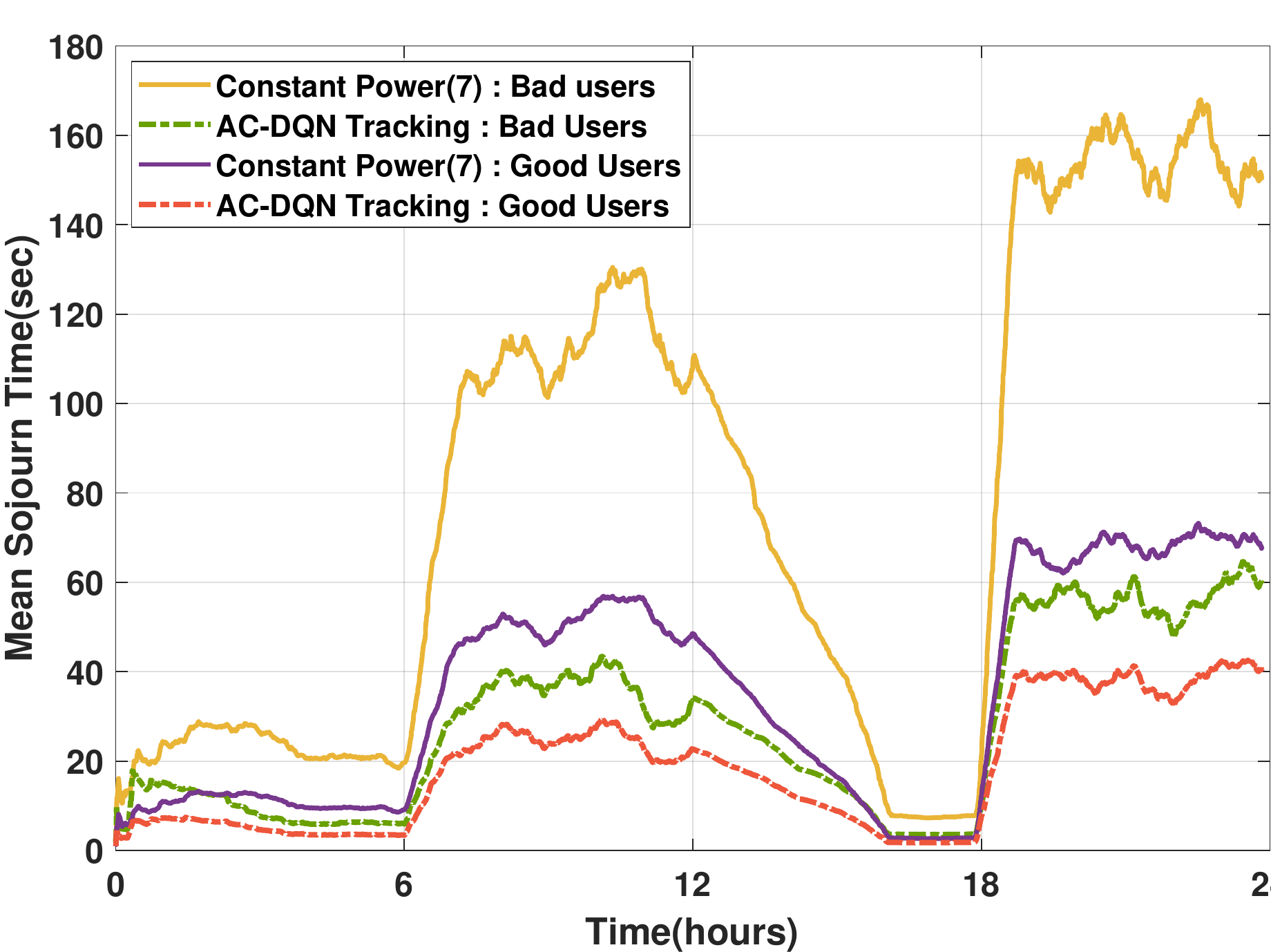}	
\caption{AC-DQN Tracking: Delay performance of AC-DQN vs Constant Power Policy, for $L=20$, $\overline{P}=7.$ }
\label{fig:track_7}
\end{figure}

\begin{figure}[h!]
\centering
	\includegraphics[trim={1.1cm 13cm 1.6cm 8cm},clip,height=3.5cm,width=8.5cm]{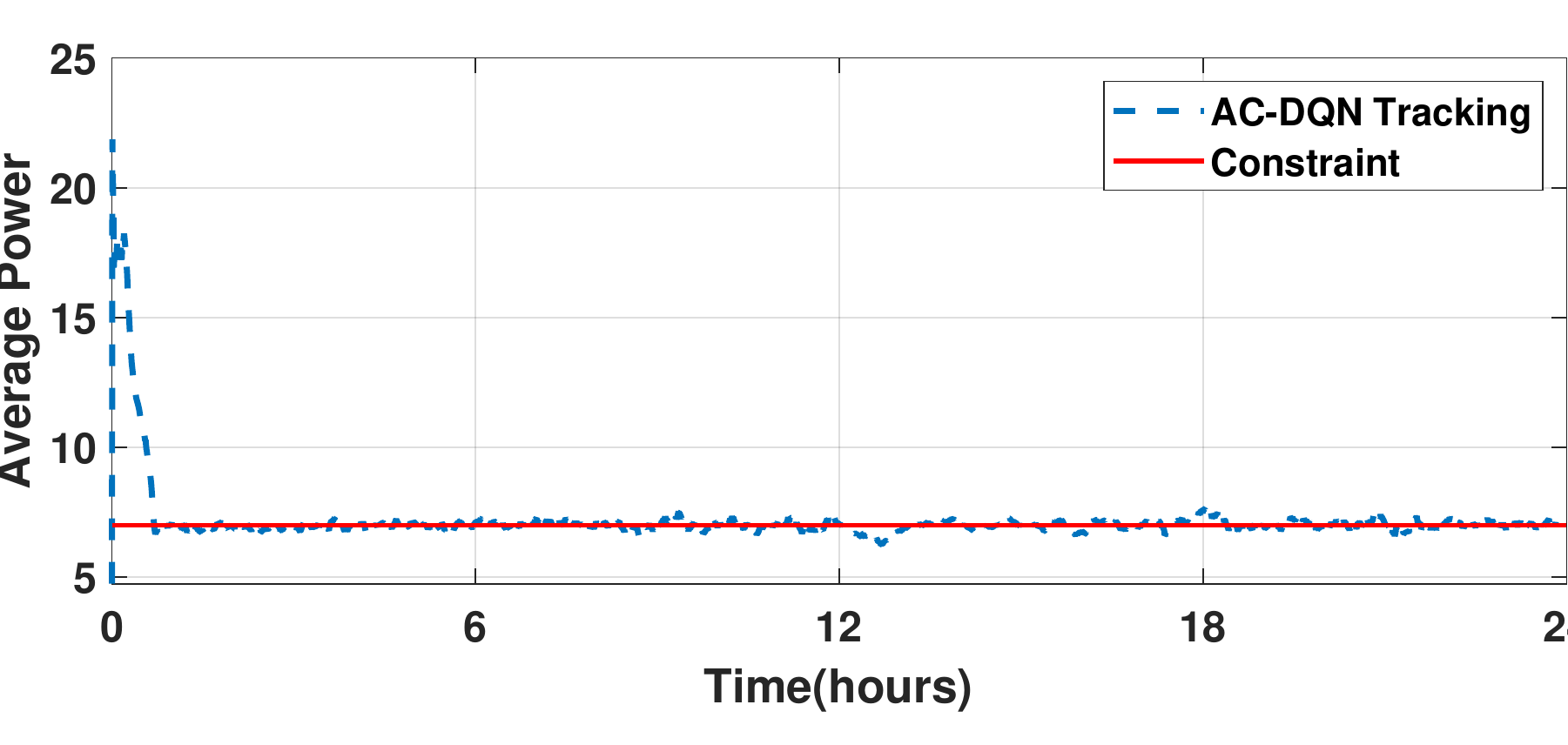}	
\caption{Tracking of Power Constraint by AC-DQN with tracking for $L=20$, $\overline{P}=7.$}	
\label{fig:track_p_7}
\end{figure}

\textbf{AC-DQN Tracking Simulations:}
In this section we show via simulations the tracking capabilities of AC-DQN. We show this for large user case with average power constraint, $\overline{P}=7$. We fix $\eta_1 = 0.001$ and $\eta_2 = 0.00003$. As explained previously, this is important for detecting the change in the environment dynamics faster. In this simulation, we vary the arrival rate at every six hours over a period of 24 hours. This captures the real world scenario where the request traffic to the base station varies with time of the day. To make the learning harder for our algorithm, we make these changes abruptly at every six hours. Specifically we use arrival rates $\lambda= 0.4, 0.8, 0.2, 1.0$ for 1st, 2nd, 3rd and 4th six hour period, respectively. We plot the AC-DQN performance for $\overline{P}=7$ in Figure \ref{fig:track_7}. We calculate the mean sojourn time and average power using a moving average window of size 1000 samples. We observe that for each arrival rate in this simulation, the AC-DQN achieves the corresponding stationary mean sojourn time performance. For instance for $\lambda=0.8$ and $\overline{P}=7$, the values in Figure \ref{fig:AC-DQN_comp_del} and Figure \ref{fig:track_7} are comparable. It is important to note that this performance is achieved while maintaining the average power constraint as can be seen in Figure \ref{fig:track_p_7}. The effect of fixing the learning rates is seen in the small oscillations of average power around $\overline{P}=7$ in Figure \ref{fig:track_p_7}. This is the oscillation in a small neighborhood around the optimal average power. Smaller the step size, lesser the oscillations. 

\begin{figure}[h!]
\includegraphics[trim={.7cm 7.6cm 1.2cm 8.1cm},clip,height=5.5cm,width=8.5cm]{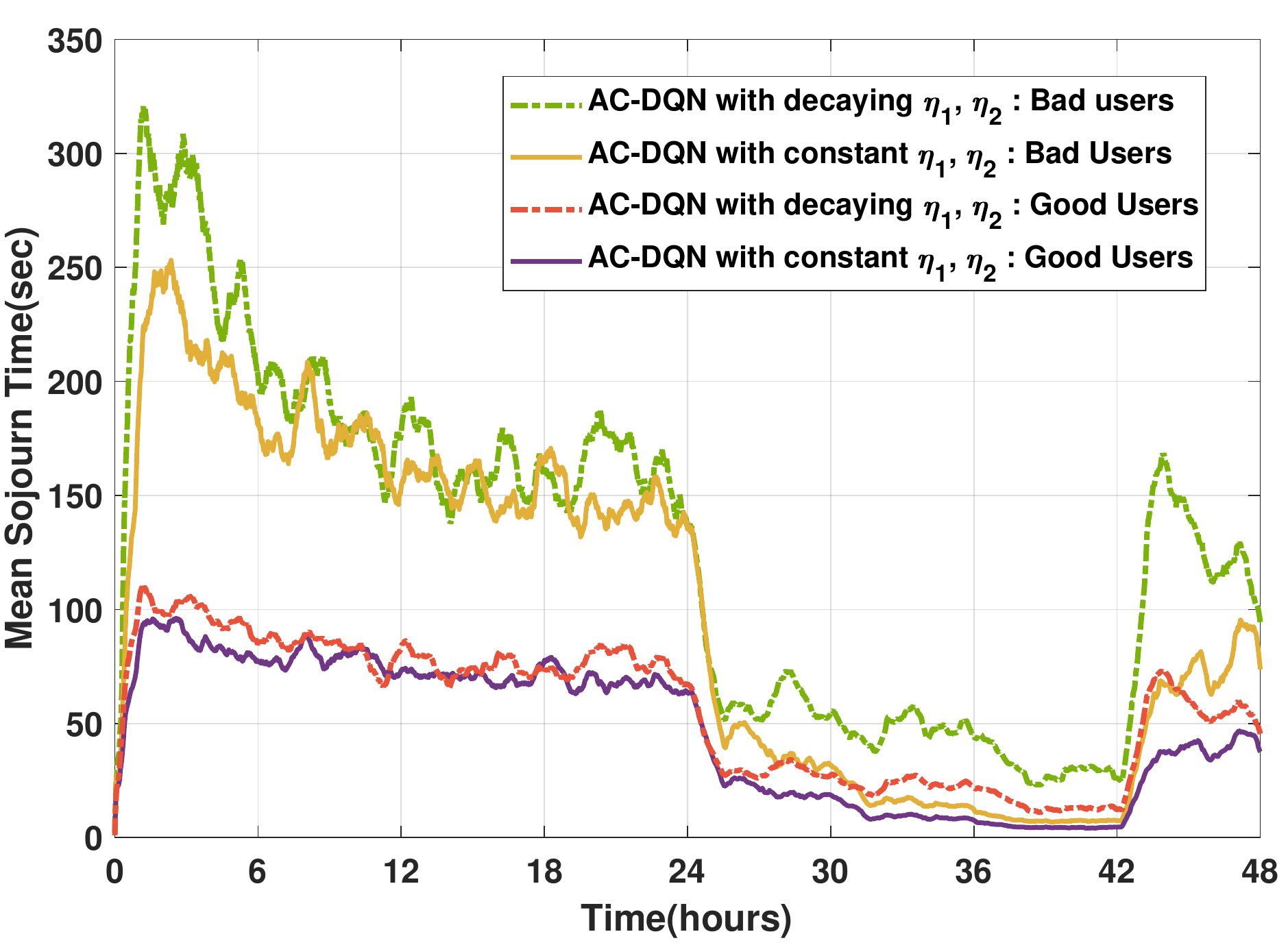}
\caption{Sojourn times for AC-DQN with decaying vs constant step-sizes. $L=20$, $\overline{P}=5$, Zipf(1) Popularity, Rayleigh fading.}
\label{fig:decay_const_del}
\end{figure}

\begin{figure}[h!]
\centering
	\includegraphics[trim={1.35cm 13cm 1.2cm 8.5cm},clip,height=3.5cm,width=8.5cm]{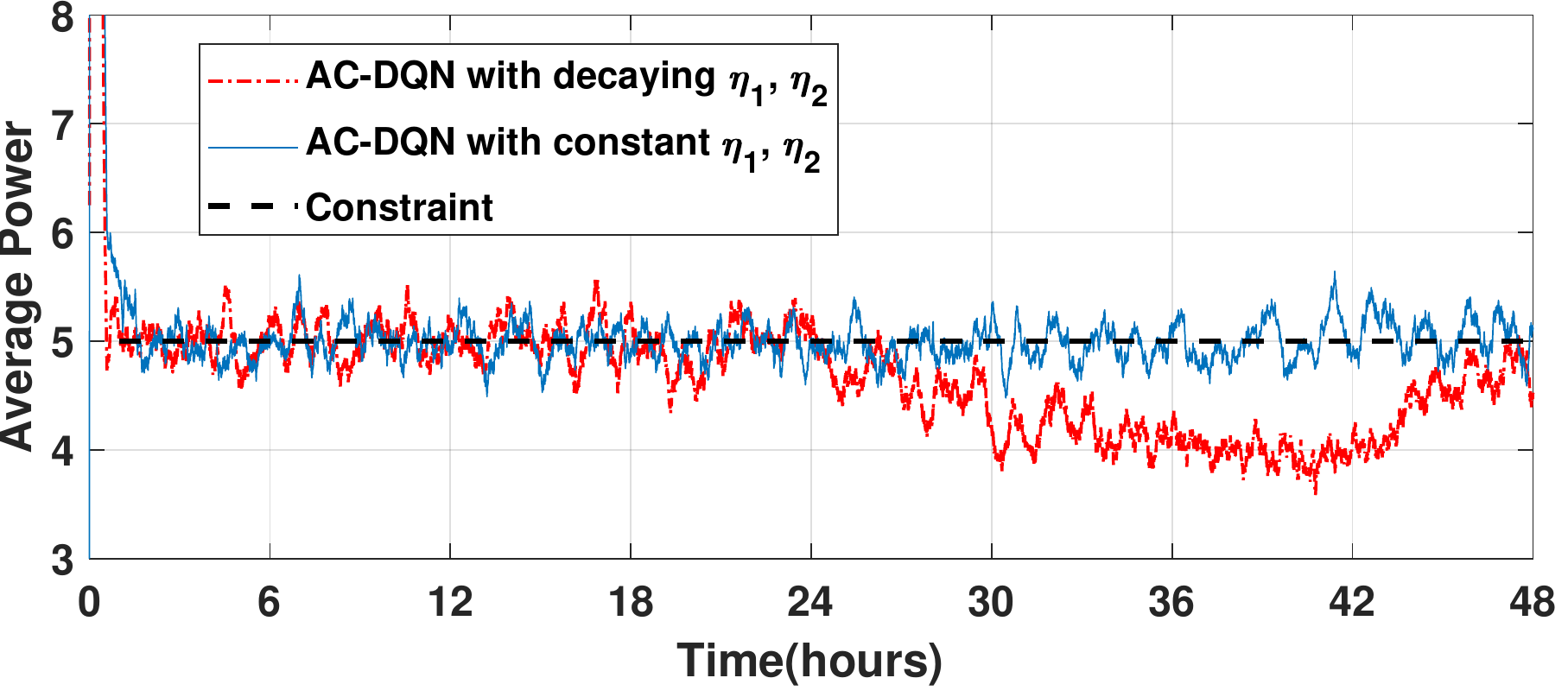}	
\caption{Convergence of Average Power for AC-DQN with decaying vs constant step-sizes. $L=20, \overline{P}=5.$}
\label{fig:decay_const_pavg}
\end{figure}

\begin{figure}[h!]
\includegraphics[trim={.7cm 7.6cm 1.2cm 8.1cm},clip,height=5.5cm,width=8.5cm]{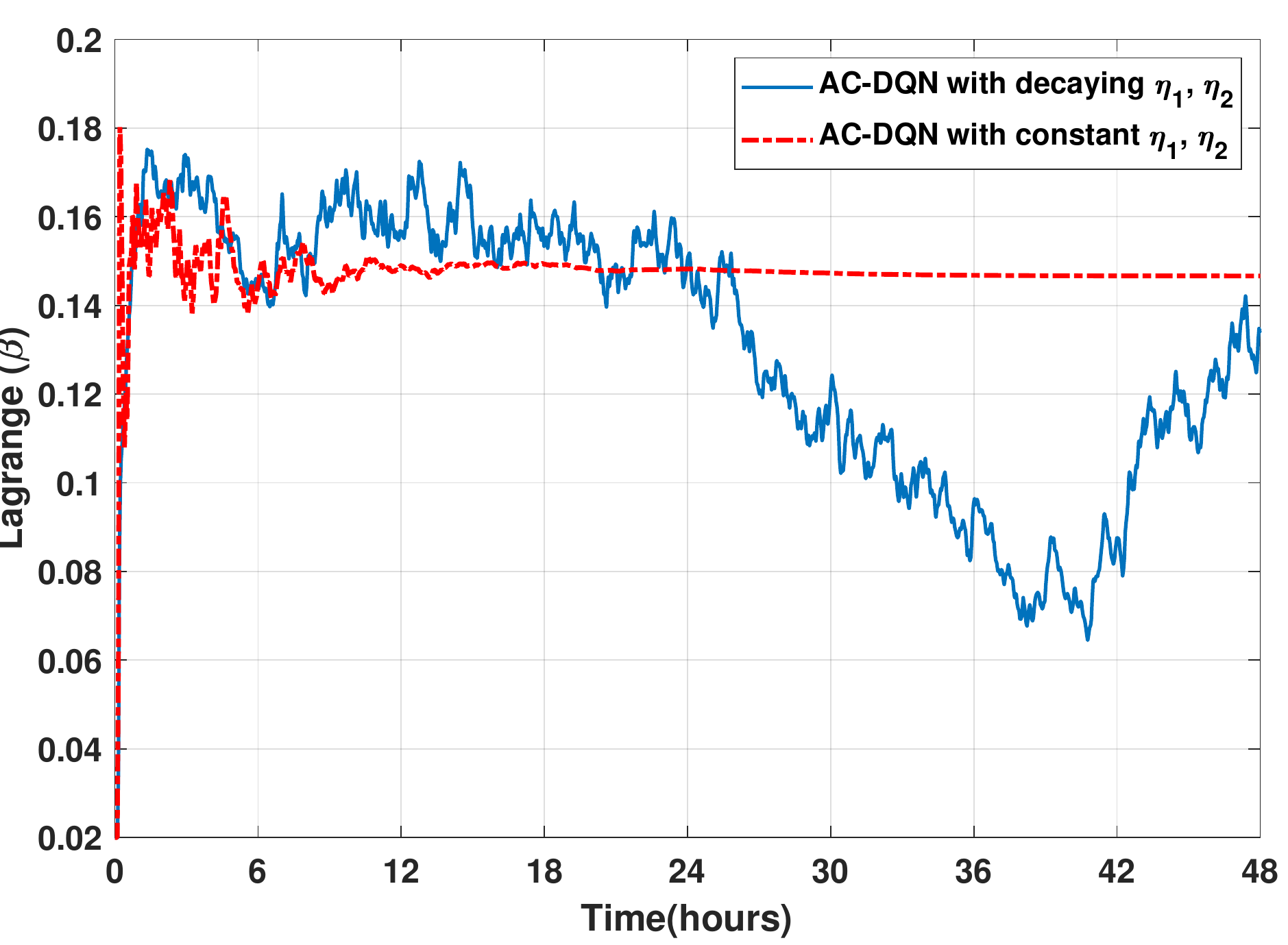}
\caption{Convergence of Lagrange for AC-DQN with decaying vs constant step-sizes. $L=20, \overline{P}=5.$}
\label{fig:decay_const_beta}
\end{figure}

Next, we demonstrate the importance of constant step sizes for $\eta_1$ and $\eta_2$, and the inability of decaying step sizes to track the changing system statistics. We consider a system where the arrival rates change over a period of 48 hours. We fix $\lambda=1.0$ for first 24 hours, then fix $\lambda=0.6, 0.5, 0.4, 0.8$ for four consecutive 6 hours intervals. This change in time period is just to illustrate the tracking ability in a more emphatic manner. It will be clear in the previous time frame also but will require more simulation time. We fix $\overline{P}=5$. We run the AC-DQN algorithm for this system with: 1) decaying $\eta_1$ and $\eta_2$ satisfying (\ref{eq:two_timeline_learning}) and 2) constant step sizes, $\eta_1=0.001$ and $\eta_2=0.00003$. Rest of the parameters remain same as in the large user case. We see in Figure \ref{fig:decay_const_del} that the AC-DQN with constant step-size almost always outperforms the decaying step size. Specifically, after the first 24 hours the delay reduction is nearly 50 percent for constant step-size. The reason for this is evident from Figures \ref{fig:decay_const_pavg} and  \ref{fig:decay_const_beta}. We see in Figure \ref{fig:decay_const_beta} that the AC-DQN with constant step-size learns the Lagrange constant through out the simulation time, whereas, the AC-DQN decaying step size is unable to learn the  Lagrange constant after the first 24 hours. As can be seen in Figure \ref{fig:decay_const_pavg}, this affects the average power achieved by the AC-DQN with decaying step size. While constant step size maintains the average power constraint of $\overline{P}=5$, the average power achieved by the decaying step-size AC-DQN drops to $4$. Hence, the decaying step-size AC-DQN suffers suboptimal utilization of available power.  Thus in practical systems only constant step-size AC-DQN will be capable of adapting to the changing system statistics.            

%

\textbf{Discussion:} We see from the simulations that the DeepRL techniques can achieve global optimal performance while providing scalability with system size. Our two-timescale approach,  AC-DQN, extends this to systems with constrained control. Though we have demonstrated this on a system with a single constraint, AC-DQN can very well be extended to systems with multiple constraints. In such systems each constraint is associated with a Lagrange constant. Each Lagrange constant adds an additional SGD step to the AC-DQN algorithm. For a stationary system, it is enough that the step-sizes satisfy multi-timescale criterion similar to (\ref{eq:two_timeline_learning}), see \cite{borkar}. However, if AC-DQN is used in systems with changing system statistics the step sizes shall be kept constant. The step sizes shall be fixed as per the tolerance requirement for a given constraint (e.g., in our system the tolerance could be $\overline{P}\pm \Delta P$. In other words, $\Delta P$ is the allowed deviation from the constraint $\overline{P}$). Lesser the tolerance, lesser the step-size. However, fixing the step-sizes too small may make the algorithm too slow to track the changes in system statistics. Hence, choosing the step sizes is a trade-off between the tolerance of the constraint and the required algorithmic agility to track the system changes.

\section{Conclusion}
\label{sec:conclusion}
We have considered a multicast downlink in a wireless network. Fading of different links to users causes significant reduction in the performance of the system. However, appropriate  change in the scheduling policies and power control can mitigate most of the losses.  However, obtaining optimal power control for this system is computationally very hard. We show that using Deep Reinforcement Learning, we can obtain optimal power control, online, even when the system statistics are unknown. We use a recently developed version  of Q learning, Deep Q Network to learn the Q-function of the system via function approximation. Furthermore, we modify the algorithm to satisfy our constraints and also to make the optimal policy track the  time varying system statistics. DDQN variant of AC-DQN provides similar performance. 

One interesting extension of this work would be adding the caches at the user nodes and learning the optimal caching policy along with the power control using DeepRL. Future works may also consider applying AC-DQN to multiple-base-station scenarios for interference mitigation. 
\bibliographystyle{IEEEtran}
\bibliography{library}
\end{document}